\documentclass[a4paper]{jpconf}
\usepackage{iopams}
\usepackage{graphicx,
longtable}

\begin{document}

\title{The Neutrinoless Double Beta Decay, Physics beyond the Standard Model and the Neutrino Mass.}
%
\author{Amand~Faessler\footnote{Talk at the Schladming  meeting, February 25th. to March 3rd. 2012.},}
\address{Institute f\"{u}r Theoretische Physik der Universit\"{a}t
T\"{u}bingen,\\
                D-72076 T\"{u}bingen, Germany}
\ead{faessler at uni-tuebingen.de}

\begin{abstract}
The Neutrinoless double beta Decay allows to determine the effectice Majorana electron neutrino mass. For this the following conditions have to be satisfied: \\(i) The neutrino must be a Majorana particle, i. e. identical to the antiparticle. \\ (ii) The half life has to be measured. \\ (iii)The transition matrix element must be reliably calculated.  \\(iv) The leading mechanism must be the light Majorana neutrino exchange. \\
The present contribution studies the accuracy with which one can calculate by different methods:(1)  Quasi-Particle Random Phase Approach (QRPA), (2) the Shell Model (SM), (3) the (before the variation) angular momentum projected Hartree-Fock-Bogoliubov method (PHFB)and the (4) Interacting Boson Approach (IBA). \\
In the second part we investigate how to determine experimentally the leading mechanism for the Neutrinoless Double Beta Decay. Is it (a)  the light Majorana neutrino exchange as one assumes to determine the effective Majorana neutrino mass, ist it the heavy left (b) or right handed (c) Majorana neutrino exchange allowed by left-right symmetric Grand Unified Theories (GUT's). Is it a mechanism due to Supersymmetry e.g. with gluino exchange and R-parity and lepton number violating terms. \\
At the end we assume, that Klapdor et al. \cite{Klapdor} have indeed measured the Neutrinoless Double Beta Decay(, although heavily contested,)and that the light Majorana neutrino exchange is the leading mechanism. With our matrix elements we obtain then an effective Majorana neutrino mass of:
\begin{eqnarray}
 <m_\nu> = 0.24 [eV](exp \pm 0.02; theor. \pm 0.01) [eV]
\end{eqnarray}
\end{abstract}


\section{Introduction}

The electron neutrino mass can be determined in the single beta decay (Mainz, Troisk and in the future KATRIN/Karlsruhe)  and for Majorana neutrinos also in the neutrinoless double beta decay. Although with both methods one has not yet succeeded to determine the effective electron neutrino mass, the neutrinoless double beta decay has in the future the potential to to measure the effective Majorana neutrino mass even to smaller values than in the single beta decay. For this the neutrino needs to be a Majorana particle (identicall with the antiparticle), the transition matrix element must be able to be calculated reliably and the light Majorana neutrino exchange must be the leading mechanisme for the neutrinoless double beta decay. \newline

The purpose of this contribution is first to investigate the accuracy, with which one can calculate the transition matrix elements. Second we want to find an experimental proceedure, which allows to determine the leading mechanisme for the neutrinoless decay: Is it the light Majorana neutrino exchange as usually assumed or the exchange of a left or right handed heavy neutrino allowed in left-right symmetric Grand Unified Theories (GUT's). An additional possible mechanisme is due to suppersymmetry (SUSY) by a trilinear term in the Lagrangian, which violates lepton number and R-parity conservation. \newline

\begin{figure}[htb]
\begin{center}
\includegraphics[scale=0.3]{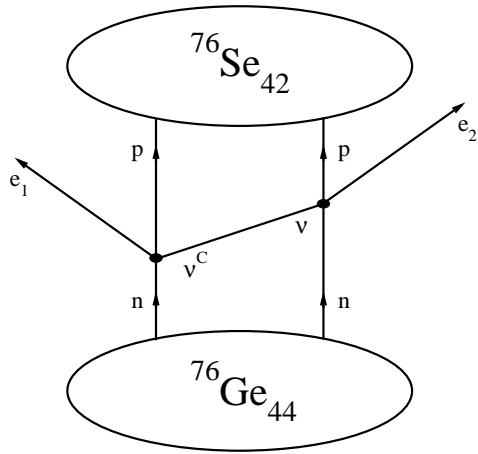}
\end{center}~~
\caption{The light Majorana neutrino exchange mechanism for the neutrinoless double beta decay of $^{76}_{32}Ge_{44}$ to $^{76}_{34}Se_{42}$ through the intermediate nucleus  $^{76}_{33}As_{43}$ is shown. the neutrino must be identicaql with the antineutrino (Majorana neutrino) and helicity must not be a good quantum number. this means the neutrino must be a massive Majorana particle.}
\label{Neutrinoless1}
\end{figure}

The transition matrix elements for the light Majorana neutrino exchange are presently calculated with four different methosd. They are reviewd in chapter, where they are compared with each other with their advantages and their drawbacks \cite{Escuderos}.  The different methods are the Quasi-particle Random Phase Approximation (QRPA) for spherical (chapter 2) and deformed nuclei  \cite{Escuderos,Rod05,Ana,Suh05,Fang1,Fang2}, the Shell Model (SM) \cite{Poves1,Poves3,Poves2}, the Projected Hartree-Fock-Bogoliubov (PHFB) approach \cite{Tomoda1, Rath1,Rath2,Rath3, Martinez} (In reference \cite{Rath3} a missing factor two in all older works of the group has been pointed out.)  and the Interacting Boson Model (IBM2)~\cite{Iachello}.
 \newline
 In the third chapter  we do not assume, that the light left handed Majorana neutrino exchange is the leading mechanism \cite{ Simkovic-Vergados, Faessler1, Faessler2}. We search for a possible experimental identification of the leading mechanism even in cases, where two or even three equally strong mechanisms interfere.
 \newline
 The diagram for the neutrinoless double beta decay with the exchange of the light Majorana neutrino is shown in figure \ref{Neutrinoless1} for the decay of $^{76}_{32}Ge_{44}$ to $^{76}_{34}Se_{42}$ through the intermediate nucleus  $^{76}_{33}As_{43}$.

  In addition to the light left handed Majorana neutrino exchange, one has other possible mechanisms as cause for the neutrinoless double beta decay: Grand Unification (GUT), Supersymmerty (SUSY) and extensions to extra dimensions. We shall discuss here extension to GUT's and SUSY.
  \newline
  In a left-right symmetric model of GUT we have:

  \begin{equation}
W_1 = \cos \vartheta_{GUT} + \sin \vartheta_{GUT} \\
W_2 = -\sin \vartheta_{GUT} + \cos \vartheta_{GUT}
\label{GUT}
\end{equation}

$W_1$ is the usual vector boson of Rubbia and coworkers of 80.4 GeV mainly responsible for the left handed weak interaction. The mixture in eq. \ref{GUT} allows at each vertex in figure \ref{Neutrinoless1} left and right handed interactions. In addition one has SUSY contributions, where mainly the threelinear terms are responsible for the lepton number violation \cite{Simkovic-Vergados, Faessler1, Faessler2, Faessler3}. This will be discussed in chapter three of this contribution.
\newline
Fermis Golden Rule of second order time dependent perturbation theory yields:
\begin{equation}
{ \cal T}^{0\nu} = \int dE_k(\nu) \sum_k \frac{<f|\hat{H}_W|k><k|\hat{H}_W|i>}{E_{0^+}(^{76}Ge) - [E_k(e^-_1) + E_k(\nu) + E_k(^{76}As)]}
\label{Golden}
\end{equation}

\begin{eqnarray}
{ \cal T}^{0\nu} =  { \cal M}^{0\nu}_\nu \cdot <m_\nu> + { \cal M}_\vartheta <\tan \vartheta> + {\cal M}_{W_R}<( \frac{M_1}{M_2})^2> + \nonumber\\ \ \ \ \ \ \ \ \ \ \ { \cal M}_{SUSY}\cdot \lambda'^2_{111} + { \cal M}_{NR}<\frac{m_p}{M_{MR}}> + ...
\label{Element}
\end{eqnarray}

The first term on the right hand side of eqn. (\ref{Element}) with the matrix element ${ \cal M}^{0\nu}_\nu$ and the effective Majorana mass

\begin{equation}
<m_\nu> = \sum_{k=1, 2, 3}(U_{ek})^2 \cdot m_{k\nu} = \sum_{k=1, 2, 3}e^{2i\alpha_k} \cdot |U_{ek}|^2 \cdot m_{k\nu}
\label{Majorana}
\end{equation}

with

\begin{equation}
\nu_e = \sum_{k=1, 2, 3} U_{ek} \nu_k
\label{neu}
\end{equation}

is the light Majorana neutrino exchange contribution.

\section{The different Many Body Approaches for the $0\nu\beta\beta$ Matrix Elements.}

  The Quasiparticle Random Phase Approximation (QRPA) is used by the groups in T\"ubingen, Bratislava and Jyv\"{a}skyl\"{a}~\cite{Escuderos,Rod05,Ana,Suh05,Fang1,Fang2}, while the Strasbourg-Madrid group ~\cite{Poves1,Poves3,Poves2} uses the Shell Model (SM), Tomoda, Faessler, Schmid and Gruemmer~\cite{Tomoda1} and Rath and coworkers~\cite{Rath1,Rath2,Rath3} use the angular momentum projected Hartree-Fock-Bogoliubov method (HFB), and Barea and Iachello~\cite{Iachello} use the Interacting Boson Model (IBM2, which distinguishes between  protons and neutrons).

The QRPA ~\cite{Escuderos,Rod05, Ana, Suh05} has the advantage to allow to use a large single-particle basis. Thus, one is able to include to each single nucleon state  in the QRPA model space also the spin-orbit partner, which guarantees that the Ikeda sum rule ~\cite{Ikeda} is fulfilled. This is essential to describe correctly the Gamow-Teller strength. The SM~\cite{Poves1} is presently still restricted to a nuclear basis of four to five single-particle levels  for the description of the neutrinoless double beta decay. Therefore, not all spin-orbit partners can be included and, as a result, the Ikeda sum rule is violated by 34 to 50\% depending on the single particle  basis used.

In QRPA one starts from the transformation to Bogoliubov quasiparticles :

\begin{equation}
a_{i}^{\dagger} = u_{i} c_{i}^{\dagger} - v_{i} c_{\bar{i}}.
\end{equation}

The creation $c_{i}^{\dagger}$ and annihilation operators of time reversed single-particle states $c_{\bar{i}}$ are usually defined with respect to oscillator wave functions~\cite{Rod05}. The single-particle energies are calculated with a Woods Saxon potential~\cite{Rod05}. \newline

The excited states $|m\rangle$ with angular momentum $J$ in the intermediate odd-odd mass nucleus are created from the correlated initial and final $0^{+}$ ground states by a proton-neutron phonon
creation operator:
\begin{equation}
|m\rangle=Q_{m}^{\dagger} |0^{+}\rangle; \ \ \hat{H} Q_{m}^{\dagger} |0^{+}\rangle = E_{m} Q_{m}^{\dagger} |0^{+}\rangle.
\label{H}
\end{equation}
\begin{equation}
Q_{m}^{\dagger} = \sum_{\alpha} [ X_{\alpha}^{m} A_{\alpha}^{\dagger} - Y_{\alpha}^{m} A_{\alpha}],
\label{Q}
\end{equation}
which is defined as a linear superposition of creation operators of proton-neutron quasiparticle pairs:
\begin{equation}
A_{\alpha}^{\dagger} = [ a_{i}^{\dagger} a_{k}^{\dagger}]_{J M},
\label{BCS0}
\end{equation}

For the present presentation the complication of angular momentum coupling, which must and is included in the quantitative calculations, is not shown.

The inverse $0\nu\beta\beta$ lifetime for the light Majorana neutrino exchange mechanism is given as the product of three factors,
\begin{equation}
\label{T1/2}
\left(T^{0\nu}_{1/2}\right)^{-1}=G^{0\nu}\,\left| { \cal M}^{0\nu}_\nu \right|^2\ \cdot <m_\nu>^2
\end{equation}
where $G^{0\nu}$ is a calculable phase space factor, $ { \cal M}^{0\nu}_\nu$ is the $0\nu\beta\beta$ nuclear
matrix element, and $<m_\nu>$ is the (nucleus-independent)
``effective Majorana neutrino mass'' (\ref{Majorana}).

The expressions for the matrix elements ${ \cal M}^{0\nu}_\nu$ and the corresponding $0\nu\beta\beta$ transition operators are given, e.g., in Ref.~\cite{Rod05}:
\begin{equation}
{ \cal M}^{(0\nu)}_\nu = { \cal M}_{GT}^{0\nu} - ( \frac{g_{V}}{g_{A}})^2 { \cal M}_F^{0\nu} -{\cal M}_T ^{0\nu}
\label{Mnu}
\end{equation}

\begin{figure}[htb]
\begin{center}
\includegraphics[scale=0.6, angle=-90
]{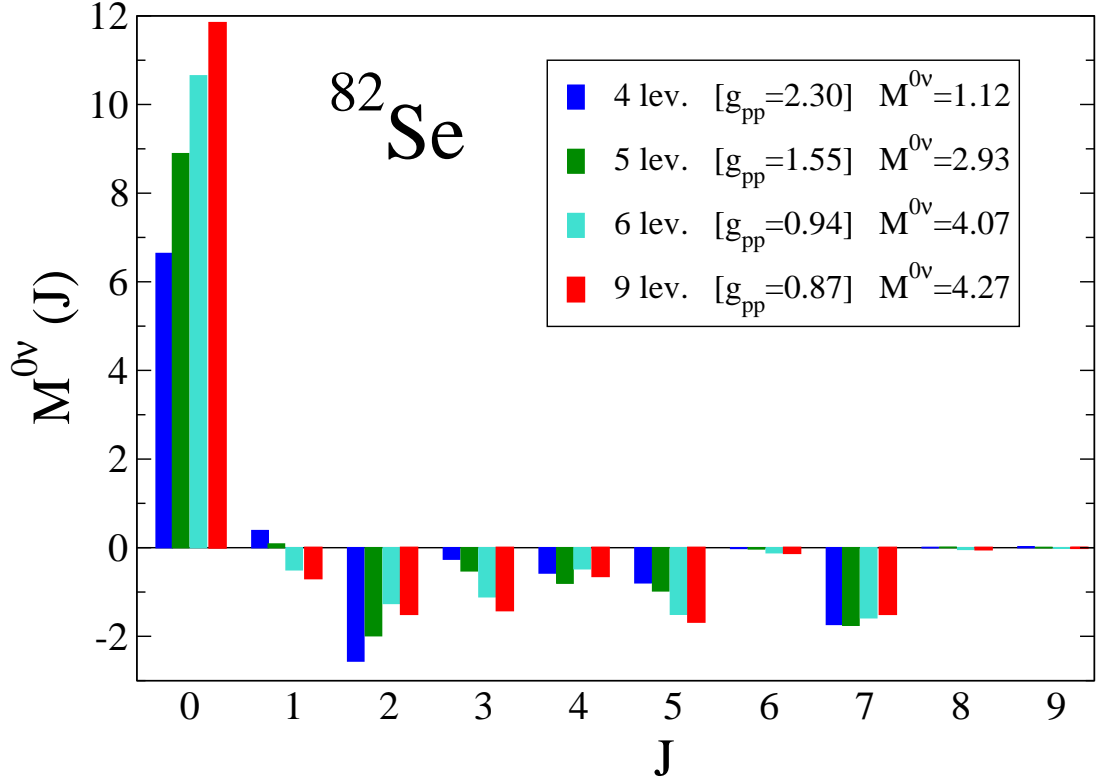}
\end{center}~~~~
\caption{(Color online) Neutron pairs with
different angular momenta $J^\pi$  transform into proton pairs with the same angular momenta. The different contributions for the angular momenta  to the total $M^{0\nu}_\nu$ calculated within the QRPA and different basis sizes for the $0\nu\beta\beta$ decay $^{82}$Se$\to^{82}$Kr is shown. The left bar is calculated with the same basis of four levels, $1p_{3/2}, 0f_{5/2}, 1p_{1/2}$ and $0g_{9/2}$, used in the shell model calculations \cite{Poves1,Poves3,Poves2}. The Ikeda Sum Rule (ISR) \cite{Ikeda} is exhausted by 50\%. The second bar from the left includes in addition the $1f_{7/2}$ level, one of the two missing spin-orbit partners given for the $^{82}$Se nucleus in ref. ~\cite{Poves3} for the shell model.  The ISR is exhausted by 66\%.  The third bar from the left  includes both missing  spin-orbit partners $0f_{7/2}$ and $0g_{7/2}$ amounting in total to 6 single-particle levels. The ISR is fulfilled by 100\%.  This leads to the increase in the neutrinoless matrix element from 1.12 to 4.07. The right bar represents the QRPA result with 9 single-particle levels $ (1f_{7/2}, 2p_{3/2}, 1f_{5/2}, 2p_{1/2}, 1g_{9/2}, 2d_{5/2}, 3s_{1/2}, 2d_{3/2}, 1g_{7/2}.)$. The matrix element gets only slightly increased to 4.27. The spin-orbit partners are essential to fulfill  the Ikeda Sum Rule (ISR).}
\label{QRPA-SM1}
\end{figure}

The SM approach has been applied by the Strasbourg-Madrid group ~\cite{Poves2} to  neutrinoless double beta decay ~\cite{Poves1}  using the 
closure relation with an averaged energy denominator. In this way one does not need to calculate the states in the odd-odd intermediate nuclei. The quality of the results depends then on the description of the $0^{+}$ ground states in the initial and final nuclei of the double beta decay system, e.g. $^{76}$Ge $\to ^{76}$Se, on the nucleon-nucleon interaction matrix elements fitted by the Oslo group in neighbouring muclei and on the average energy denominator chosen (fitted) for closure. The $0\nu\beta\beta$ transition matrix element (\ref{Mnu}) simplifies as shown in  equations (5) to (11) of  Ref.~\cite{Poves3}. Since the number of many body configurations is increasing drastically with the single-particle basis, one is forced to restrict for mass numbers $A = 76$ and $A = 82$ in the SM the single-particle basis to $1p_{3/2}, 0f_{5/2}, 1p_{1/2}$ and $0g_{9/2}$. In ref.~\cite{Poves3} the $^{82}$Se nucleus is calculated in the SM for five basis single-particle levels including also $0f_{7/2}$. For the mass region around $A = 130$ the SM basis is restricted to $0g_{7/2}, 1d_{3/2}, 1d_{5/2}, 2s_{1/2}$  and  $0h_{11/2}$ levels. The problem with these small basis sets is that the spin-orbit partners $0f_{7/2}$ and $0g_{7/2}$ have to be omitted~\cite{Poves3}. The SM results then automatically violate the Ikeda Sum Rule (ISR)~\cite{Ikeda}, while the QRPA satisfies it exactly. The Ikeda sum rule is:
\begin{equation}
S_- - S_+ = 3(N - Z),
\label{Ikeda}
\end{equation}
\begin{equation}
S_- = \sum_{\mu} \langle 0^{+}_i|[ \sum_{k}^{A} (-)^{\mu}\sigma_{-\mu}(k) t_{+}(k)] [ \sum_{l}^{A} \sigma_{\mu}(l) t_{-}(l)] |0^{+}_i \rangle,
\label{S}
\end{equation}
For $S_{+}$ the subscripts at the isospin rising and lowering operators are exchanged.

Figure~\ref{QRPA-SM1}  shows the QRPA contributions of different angular momenta of the neutron pairs, which are changed in  proton pairs with the same angular momenta. In figure ~\ref{QRPA-SM1} the left bar is the result for $^{82}$Se obtained with the single-particle basis $1p_{3/2}, 0f_{5/2}, 1p_{1/2}$ and $0g_{9/2}$ used in the SM.  The ISR is exhausted by 50\% only. The second bar from the left represents the result with addition of the $1f_{7/2}$ level. The ISR is exhausted by 66\%. The third bar from the left shows the result obtained by inclusion of both  spin-orbit partners $0f_{7/2}$ and $0g_{7/2}$  missing in the four level basis of the SM. The ISR is 100\% fulfilled. For the right bar the basis is increased to 9 single-particle levels for neutrons and protons ($ 0f_{7/2}, 1p_{3/2}, 0f_{5/2}, 1p_{1/2}, 0g_{9/2}, 1d_{5/2}, 2s_{1/2}, 1d_{3/2}, 0g_{7/2}$).

In the last ten years P. K. Rath and coworkers ~\cite{Rath1, Rath2, Rath3} have published a whole series of papers (see references in Ref.~\cite{Rath2}) on $2\nu\beta\beta$ decay and, since 2008, also on $0\nu\beta\beta$ decay, in which they used a simple pairing plus quadrupole many body Hamiltonian of the Kumar and Baranger type~\cite{Kumar} to calculate the neutrinoless double beta decay transition matrix elements with angular momentum projection from a Hartree-Fock-Bogoliubov (HFB) wave function after variation. Schmid ~\cite{Schmid2} did show, that with the assumption of a real Bogoliubov transformation (real coefficient A and B), axial symmetry

\begin{equation}
a_{\alpha}^{\dagger} = \sum_{i = 1}^M (A_{i \alpha}c_{i}^{\dagger}  + B_{i \alpha} c_i)
\label{Bogo}
\end{equation}
and no parity mixing, only $0^+, 2^+, 4^+, \dots.$ nucleon pairs and excited states are allowed (See eqn. (4.2.3) on page 603 of ref. \cite{Schmid2}).   Rodriguez and Martinez-Pinedo start with the projected HFB approach but allow admixtures of different deformations using the Generator Coordinate Method (GCM) and the Gogny force \cite{Gogny}.
The QRPA and the SM do not have this restriction like PHFB and also in the PHFB with the deformation GCM extension (GCM-PNAMP) \cite{Martinez}.

\begin{figure}[htb]
\begin{center}

\

\

\

\

\includegraphics[scale=0.65
]{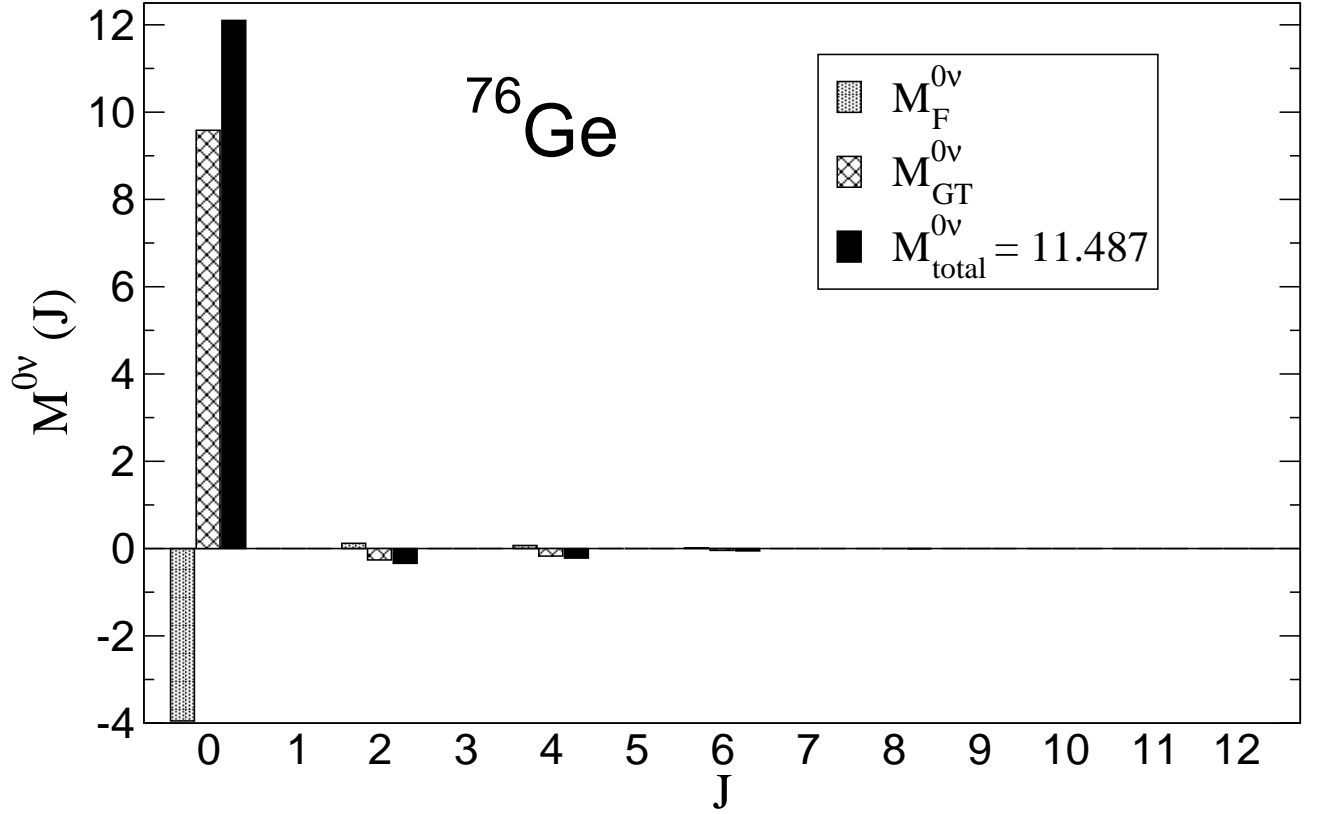}
\end{center}~~
\caption{
Neutron pairs with different angular momenta contributing  to the neutrinoless double beta decay transition matrix elements for $^{76}$Ge $\to ^{76}$Se calculated from a HFB wave function with angular momentum and proton and neutron particle number projection before variation. The Fermi, the Gamow-Teller and the total contribution including the tensor part as defined in eq.~(\ref{Mnu}) are separately given. The nucleon-nucleon interaction is an improved Gogny type force~\cite{Gogny}. The deformations $\beta_{Ge} =-0.08$ and $\beta_{Se} = 0.11$ correspond to the minima of the projected HFB total energy. The results for the transition matrix elements are qualitatively and almost quantitatively the same for the experimental deformation from the Coulomb reorientation effect: $ \beta_{Ge} = 0.16, \beta_{Se} = 0.10 $ and also for different forces. The angular momenta of the  neutron pairs are in the PHFB approach with axial symmetry, real coefficients and no parity mixing  restricted  to $0^+, 2^+, 4^+, \dots $. In addition the contributions of higher angular momentum neutron pairs $2^+, 4^+, \dots $ are drastically reduced compared to the QRPA and the SM. }
\label{HFB-Fig}
\end{figure}

Figure \ref{HFB-Fig} shows on the other side, that the projected HFB approach is restricted to contributions of neutron pairs with angular momenta $0^+, 2^+, 4^+, \dots $. In addition, one sees that the contributions of transition of higher angular momentum neutron to proton pairs $2^+, 4^+, \dots$  are drastically reduced compared to the QRPA and the SM see fig. \ref{HFB-Fig}. The reason for this is obvious: in a spherical nucleus the HFB solution contains only seniority zero and no stronger higher angular momentum pairs. The double beta decay system $^{76}_{32}Ge_{44} \to^{76}_{34}Se_{42}$ has only small deformations and thus a projected HFB state is not able to describe an appreciable admixture of higher angular momentum pairs for $0^+ \to 0^+$ transitions as can be seen in ref. \cite{Schmid2}. The higher angular momentum contributions increase drastically with increasing intrinsic quadrupole and hexadecapole deformations of the HFB solution. \\

The results in figure ~\ref{HFB-Fig} are calculated in.Tuebingen by K.W. Schmid~\cite{Schmid2} within the HFB with angular momentum and particle number projection before variation with an improved Gogny force~\cite{Gogny} adjusted in a global fit to properties of many nuclei.

To have also $1^+, 3^+, 5^+, \dots $ neutron pairs contributing one has to use a Bogoliubov transformation  with complex coefficients A and B (\ref{Bogo}). To have also $0^-, 1^-, 2^-, 3^-, 4^-, 5^-, \dots $ one has to allow parity mixing in the Bogoliubov transformation (\ref{Bogo}). But even allowing all different types of angular momentum and parity pairs one would still have an unnatural suppression of the higher angular momenta especially for smaller deformations. This handicap could probably be overcome by a multi-configuration HFB wave function \cite{Schmid2} with complex coefficients and parity mixing in the Bogoliubov (\ref{Bogo}) transformation.

The IBM (Interacting Boson Model)~\cite{Iachello} can only change $0^+$ (S) and $2^+ $ (D) fermionic pairs from two neutrons into two protons. In the bosonization to higher orders this leads to the creation and annihilation of up to three ``s'' and ``d'' boson annihilation and creation operators in Ref.~\cite{Iachello}. But all these terms  of equation (18) of reference~\cite{Iachello} originate from the annihilation of a $0^+$ (S) or a $2^+ $ (D) neutron pair into a corresponding proton pair with the same angular momentum. The higher boson terms try only to fulfill the Fermi commutation relations of the original nucleon pairs up to third order. The IBM can therefore change only a $0^+$ or a $2^+$ neutron pair into a corresponding proton pair.

\begin{table}[h]
\centering
\caption{Nuclear matrix elements (NME) for deforemed "1" and for spherical "0"  QRPA calculations  $ { \cal M}^{0\nu}_\nu$ for $0\nu\beta\beta$ decays $^{76}$Ge$\rightarrow ^{76}$Se, $^{150}$Nd$\rightarrow ^{150}$Sm, $^{160}$Gd$\rightarrow ^{160}Dy$. The BCS overlaps  are taken into account.  In the last two columns the $0\nu\beta\beta$ matrix element ${ \cal M}^{0\nu}_\nu $ and the half-lives for assumed $ <m_\nu>= 50 $ meV are shown.
}
\label{table.2}
\begin{tabular}{|c | c| c | c| c|}
	\hline
& & & & \\
$A$ & Def. &$g_A$ &${M}^{0\nu}_\nu$& $T^{0\nu}_{1/2} \cdot [10^{26}y] $
\\[2pt]
& & & & $<m_\nu>$=50 meV
\\[4pt]	
\hline
76 & ``1" & 1.25 & 4.69
& 7.15\\
 & ``0" & 1.25
& 5.30
& 5.60\\

\hline
\hline
150 & ``1" & 1.25
& 3.34
& 0.41 \\
 & ``0" & 1.25
& 6.12
& 0.12\\
\hline
\hline
160 & ``1"  & 1.25
& 3.76
& 2.26\\
\hline
\end{tabular}
\label{MEres}
\end{table}

We have also calculated \cite{Fang1,Fang2} the transition matrix elements of the light left handed Majorana neutrino exchange ${ \cal M}^{0\nu}_\nu $ including deformations. Different deformations for the inital and the final nuclei are allowed. The BCS overlaps are included. The quadrupole deformations are taken from the reorientation Coulomb excitation of the $2^+$ states.

 \begin{equation}
\beta_2 = \sqrt{ \frac{\pi}{5}} \frac{Q_{reorientation}}{Z<r^2>_{charge}}
\label{beta}
\end{equation}

\begin{figure}[t]
\begin{center}

\

\

\

\

\includegraphics[scale=0.6
]{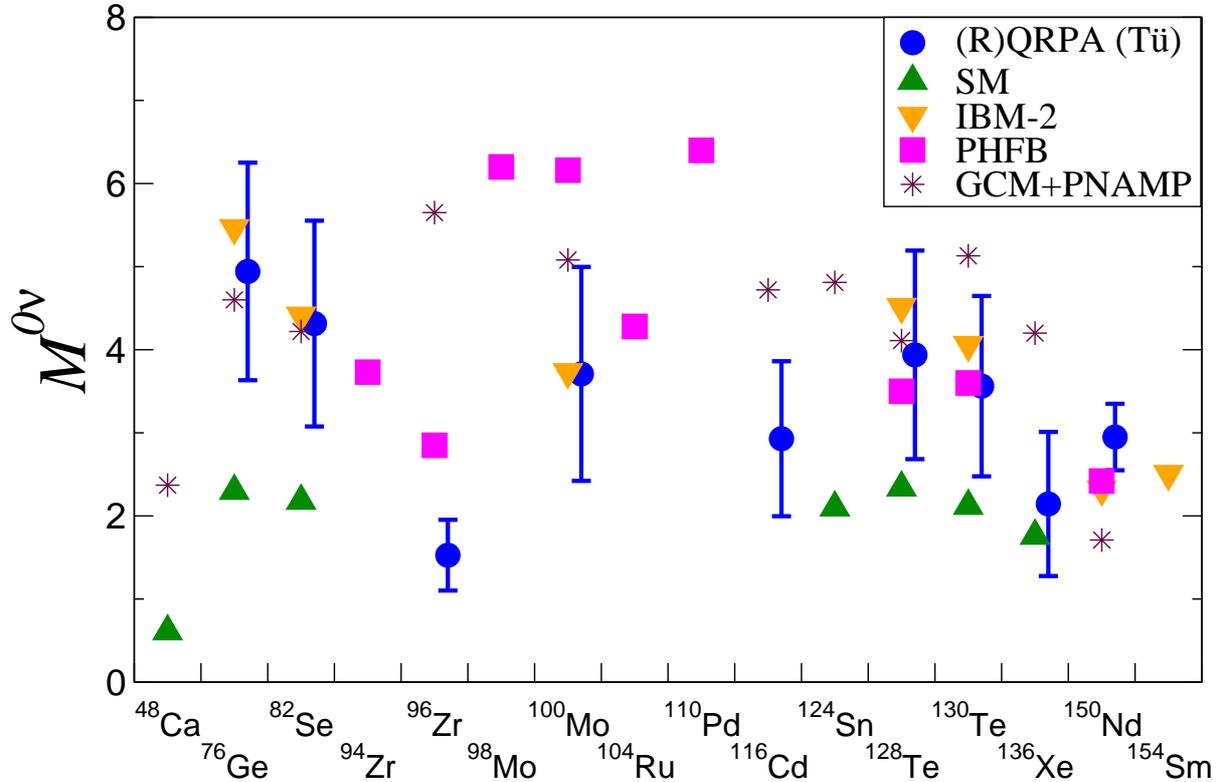}
\end{center}
\caption{(Color online) \ Transition matrix elements of the neutrinoless double beta decay  for the different approaches: QRPA~\cite{Rod05, Ana}, the SM 
\cite{Poves1, Poves2, Poves3}, the projected HFB 
method~\cite{Rath3}, the projected HFB with the Generator Coordinate Method (GCM) with deformations \cite{Martinez} (GCM-PNAMP) and the IBM2
\cite{Iachello}. The error bars of the filled circles  for the QRPA are calculated as the highest and the lowest values for three different single-particle basis sets, two forces (Bonn CD and Argonne V18)
two different axial charges $g_A = 1.25$ and the quenched value $g_A = 1.00$ and two different treatments of short range correlations (Jastrow-like~\cite{Miller} and the Unitary Correlator  Operator Method (UCOM)~\cite{Feld}). The radius parameter is as in this whole work $r_{0} = 1.2$ fm. The triangle with the tip up are the SM results \cite{Poves1, Poves2, Poves3}. The triangle with the tip down represent the transition matrix element of the Interacting Boson Model 2 (IBM2) \cite{Iachello}. The squares have been calculated by Pradfulla Rath and coworkers \cite {Rath3} with the correction of the factor 2  from December 2010 included, with which all previous results of Rath et al. have to be multiplied \cite{Rath1,Rath2,Rath3}. The star (GCM-PNAMP) is a projected HFB calculation with  the Gogny force \cite{Gogny} by Rodriguez and Martinez-Pinedo \cite{Martinez} allowing for different deformations with the Generator Coordinate Method (GCM).}
\label{total}
\end{figure}

The deformation reduces the matrix elements \cite{Fang1,Fang2} in $^{76}Ge$ slightly from $5.30$ to $4.69$ by $10 \%$ only. This is within the error of the matrix elements (see figure \ref{total}). But the reduction of the matrix elements is severe in strongly deformed systems with different deformations for the initial and the final nuclei. In the system $^{150}Nd \to ^{150}Sm$ the matrix element is reduced from $6.12 $ to $3.34$  (see table 2) and in the strongly deformed system $^{160}Gd \to ^{160}Dy$ (see table 1) one obtains a matrix element of $3.76$ (see table 2). The single nucleon basis in these deformed calculations are determined in a deformed Woods-Saxon potential. The results are then expanded into a deformed oscillator basis with the same deformation parameter and the appropriate oscillator length in seven oscillator shells \cite{Fang1,Fang2}. The deformed result for  $^{150}Nd \to ^{150}Sm$ is  included in figure \ref{total}.

\section{How to find the Leading  Mechanisms for the Neutrinoless Double Beta Decay?}

Normally one assumes, that the first term of eq. (\ref{Element}) is the leading one and with the experimental data and the matrix element for the light left handed Majorana neutrino exchange $ {\cal M}^{0\nu}_\nu$ one can determine the effective Majorana neutrino mass (\ref{Majorana}). But in Grand Unification (GUT) and Supersymmetry (SUSY) additional mechanisms for the neutrinoless Double Beta Decay $(0\nu\beta\beta)$ are possible.

\begin{figure}[htb]
\begin{center}
\includegraphics[scale=0.3, angle=-90]{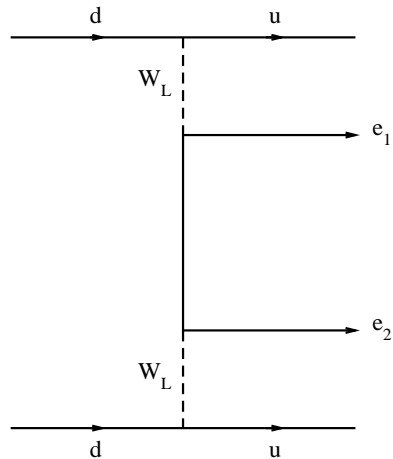}
\end{center}~~
\caption{Light Majorana neutrino exchange. The two neutrons in the initial nucleus in the ground state $0^+$, which change into two protons are are coming from the left and are characterized by the two down quarks d  in the two neutrons, which change into two up quarks u  in two protons in the final nucleus. The vector bosons $W_L$ mediating the left handed weak interaction are coupled by the expansion coefficients $U_{ek}$ ( \ref{neu}) to the neutrino mass eigenstates $m_{k\nu}$. }
\label{Neutrinoless2}
\end{figure}

The matrix element of this gold plated term is  proportional to:

\begin{eqnarray}
{ \cal M}^{0\nu}_\nu(light \ \nu_L) \propto \sum_{k=1,2,3} U^\nu_{ek}\cdot P_L \frac{1}{\not\!q -m_{k\nu}} P_L \cdot U^\nu_{ek} = \frac{1}{q^2} \sum_{k=1, 2, 3} e^{2i\alpha^\nu_k} \cdot |U^\nu_{ek}|^2 m_{k\nu}
\label{Element1}
\end{eqnarray}
The exchange of a heavy left handed Majorana neutrino:

\begin{equation}
N_e = \sum_{k=1, ... 6} U^N_{ek} N_k \approx \sum_{k=4,5,6} e^{i\alpha^N_k} |U^N_{ek}|\cdot N_{k}
\label{nuh}
\end{equation}

with $\alpha^N_k$ the Majorana phases for these heavy left handed Majorana neutrinos.

\begin{eqnarray}
{ \cal M}^{0\nu}(heavy \ N_L) \propto \sum_{k=4,5,6} U^{N}_{ek}\cdot P_L \frac{1}{\not\!q -M_{kN}} P_L \cdot U^{N}_{ek} = -  \sum_{k=4,5,6}e^{2i\alpha^N_k} \cdot  |U^N_{ek}|^2/M_{kN}
\label{Element2}
\end{eqnarray}

The lepton number and R-parity  violating contributions in SUSY are the trilinear terms and the coupling of the lepton superfields to the Higgs particle.

\begin{equation}
W_{\not\!R} = \lambda_{ijk} \cdot L_i \cdot L_j \cdot E^c_k + \lambda'_{ijk} \cdot L_i\cdot Q_j \cdot D^c_k + \mu_i\cdot L_i\cdot H_2
\label{tri}
\end{equation}
The lepton L, E  and quark Q, D  left (L) and right (R) handed superfields are defined as:

\begin{eqnarray}
L_k = \left( \matrix{ \nu \cr e\cr \tilde{\nu} \cr \tilde{e}}\right)_{kL}; \  E_k = \left( \matrix{ e \cr \tilde{e}}\right)_{kR};\  Q_k = \left( \matrix{ u \cr d\cr \tilde{u} \cr \tilde{d}}\right)_{kL};\ D_k = \left( \matrix{ d \cr \tilde{d}}\right)_{kR};
\label{super}
\end{eqnarray}
The indices i,j,k run over the three families for leptons: $ e, \mu, \tau $ and for quarks: $ d, s, b $. The subscripts L and R characterize left and right handed superfields. The tilde indicates SUSY particles like selectrons , sneutrinos and squarks.

The inverse half life is given by:

\begin{eqnarray}
\frac{1}{T^{0\nu}_{1/2}} = \frac{w^{0\nu}}{\ln 2} \approx G^{0\nu}(E_0,Z) \ \cdot |[ \eta_\nu { \cal M}^{0\nu}_\nu + \nonumber\\  \eta_{NL}{ \cal M }^{0\nu}_{NL} + \eta_{\lambda'} { \cal M}^{0\nu}_{\lambda'}]^2 +  |\eta_{NR} |^2 | { \cal M}^{0\nu}_{NR}|^2 |
\label{trans}
\end{eqnarray}
with:
\begin{eqnarray}
\eta_\nu =\frac{<m_\nu>}{m_e} =   ( |U^\nu_{e1}|^2\cdot m_1 + e^{2i\alpha_{21}} \cdot |U^\nu_{e2}|^2 \cdot m_2 + e^{2i\alpha_{31}}\cdot |U^\nu_{e3}|^2 \cdot m_3)/m_e
\label{lightnu}
\end{eqnarray}

\begin{eqnarray}
\eta_{NL} =  |U^N_{e4}|^2\cdot\frac{m_p}{M_{4L}} +   e^{2i\alpha_{54}} \cdot |U^N_{e5}|^2 \cdot \frac{m_p}{M_{5L}}  + e^{2i\alpha_{64}}\cdot |U^N_{e6}|^2 \cdot \frac{m_p}{M_{6L}}
\label{heavynu}
\end{eqnarray}

\begin{eqnarray}
\alpha_{ik} = \alpha_i - \alpha_k
\label{phase}
\end{eqnarray}

We restrict here for SUSY to the trilinear terms (\ref{tri}).They are lepton number and R parity violating. They can contribute by gluino or by neutralino exchange.\newline
To test, if the light left handed Majorana neutrino exchange is the leading mechanism, one meeds at least experimental data of the neutrinoless double beta decay and reliable transition matrix elements in two systems.  The light Majorana neutrino exchange is represented by the gold plated term, which is the first on the right hand side of eq.(\ref{Element}). If the measurements and the matrix element are reliable enough \cite{Faessler2} and the light Majorana neutrino exchange is indeed the leading mechanism, one should in both and also in all other systems obtain the same effective Majorana neutrino mass. If two mechanisms are at the same time contributing, one must distinguish between non-interfering and between interfering mechanisms. The light left handed Majorana  neutrino  and the heavy right handed neutrino exchange have negligible interference \cite{Faessler1}.

\begin{eqnarray}
\frac{1}{T^{0\nu}_{i,1/2}\cdot G^{0\nu}_i(E_0,Z)} = | \eta_\nu|^2( { \cal M}^{0\nu}_{i,\nu})^2  +  |\eta_{NR} |^2 ( { \cal M}^{0\nu}_{i,NR})^2
\label{transno}
\end{eqnarray}

To determine the absolute values of the two strength parameters $\eta_\nu$ and $ \eta_{NR}$ one needs at least two decay systems i. To verify, that these are indeed the leading mechanisms one needs at least a measurement in one additional decay system i. But if one forms ratios of half lives using the Tuebingen matrix elements for the light Majorana neutrino exchange and the heavy right handed neutrino exchange one otaines a very restricted allowed interval for these ratios \cite{Faessler1}.

\begin{eqnarray}
0.15 \le \frac{T^{0\nu}_{1/2}(^{100}Mo)}{T^{0\nu}_{1/2}(^{76}Ge)} \le 0.18; \ \ \ \ \ \ \ \
0.17 \le \frac{T^{0\nu}_{1/2}(^{130}Te)}{T^{0\nu}_{1/2}(^{76}Ge)} \le 0.22; \nonumber\\
1.14 \le \frac{T^{0\nu}_{1/2}(^{130}Te)}{T^{0\nu}_{1/2}(^{100}Mo)} \le 1.24;
\label{ratios1}
\end{eqnarray}

The dependence of this ratios on the different parameters of the nuclear structure calculation is very minor. Due to the ratios the dependence on most changes drop approximately out. The ratios (\ref{ratios1}) are calculated for the axial charge $ g_A = 1.25$. But the quenching of this value to $g_A = 1.00$ has only a minor effect \cite{Faessler1}.
\newline
If the two leading mechanisms like the light Majorana neutrino exchange and the SUSY mechanism with gluino or neutralino exchange can interfer, the situation is a bit more complicated: Let us assume the relative phase angle of the complex strength parameters $\eta_{\nu} $ and $\eta_{\lambda'}$ is $ \vartheta_{\nu,\lambda'}$. The inverse of the half life time the phase space factor $ G^{0\nu}_{1/2}(E_0,Z)$ is then:

\begin{eqnarray}
\frac{1}{T^{0\nu}_{i,1/2}\cdot G^{0\nu}_i(E_0,Z)} = | \eta_\nu|^2( { \cal M}^{0\nu}_{i,\nu})^2  +  |\eta_{\lambda'} |^2 ( { \cal M}^{0\nu}_{i,\lambda'})^2 + \nonumber\\
 \cos \vartheta_{\nu,\lambda'} \cdot |\eta_\nu| \cdot |\eta_{\lambda'}| \cdot
{\cal M}^{0\nu}_{i,\nu} \cdot { \cal M}^{0\nu}_{i,\lambda'};
\label{transint}
\end{eqnarray}

One needs three decay systems to determine the absolute values of the parameters $ \eta_\nu, \eta_{\lambda'}$ and the relative phase angle $\vartheta_{\nu, \lambda'} $. At least one additional system is needed to verify, that indeed these two mechanisms are the leading ones. Again the ratios of the half lives are allowed to lie only in narrow regions \cite{Faessler1}. If this is not the case, the chosen mechanisms are not the leading ones \cite{Faessler1}. With CP conservation the strength parameters $\eta$ must be real and thus the relative phase angle is zero or 180 degrees. So the determination of  $\vartheta_{\nu,\lambda'}$  allows to test CP conservation or violation.

\section{ The effective Majorana Neutrino Mass.}
Before we summarize the results let us assume Klapdor-Kleingrothaus et al. \cite{Klapdor} have indeed measured the neutrinoless double beta decay in $^{76}Ge$, although the general belief is, that this still needs confirmation. From the half life given by Klapdor et al. \cite{Klapdor} one can derive with our matrix elements the effective Majorana neutrino mass (\ref{Majorana}).

\begin{eqnarray}
T^{0\nu}_{1/2}(^{76}Ge, exp\ Klapdor) = (2.23 + 0.44 -0.31)\cdot 10^{25} [years];
\label{Klapdor}
\end{eqnarray}

With our matrix elements one obtains the effective light left handed  Majorana neutrino mass under the assumption, that the light Majorana exchange is the leading mechanism.

 \begin{eqnarray}
 <m_\nu> = 0.24 [eV](exp \pm 0.02; theor. \pm 0.01) [eV]
\label{theory}
\end{eqnarray}

The uncertainty (error) from experiment is 0.02 [eV], while the theoretical  error originates from the uncertainties of the QRPA matrix elements as indicated in figure \ref{total}. The theoretical error is 0.01 [eV].

\section{Conclusions}

Let us now summarize the results of this contribution:

The Shell Model (SM) \cite{Poves1, Poves2, Poves3} is in principle the best method to calculate the nuclear matrix elements for the neutrinoless double beta decay. But due to the restricted single-particle basis it has a severe handicap. The matrix elements in the $^{76}Ge$ region are by a factor 2  smaller than the results of the Quasiparticle Random Phase Approximation (QRPA)~\cite{Escuderos,Rod05,Ana,Suh05}, the projected Hartree Fock Bogoliubov approach \cite{Rath3, Martinez} and the Interacting Boson Model (IBM2) \cite{Iachello}. With the same restricted basis as used by the SM the QRPA obtains roughly the same results as the SM (figure \ref{QRPA-SM1}), but the Ikeda sum rule~\cite{Ikeda} gets strongly violated due to the missing  spin-orbit partners in the SM single-particle basis.

The angular momentum projected Hartee-Fock-Bogoliubov (HFB) method \cite{Rath1} is restricted in its scope. With a real Bogoliubov transformation without parity mixing and with  axial symmetry (\ref{Bogo}) one can only describe neutron pairs with angular momenta and parity $ 0^+, 2^+, 4^+, 6^+, \dots $ changing into two protons for ground state-to-ground state transitions. The restriction for the Interacting Boson Model (IBM) \cite{Iachello} is even more severe: one is restricted to $0^+$ and $2^+$ neutron pairs changing into two protons.

A comparison of the $0\nu\beta\beta$ transition matrix elements calculated recently in the different many body methods: QRPA with realistic forces (CD Bonn, Argonne V18), SM with nucleon-nucleon matrix elements fitted in neughbouring nuclei, projected HFB \cite{Rath3} with pairing plus quadrupole force \cite{Kumar}, projected HFB with the deformation as Generator Coordinate (GCM+PNAMP)\cite{Martinez} and with the Gogny force\cite{Gogny} and IBM2 \cite{ Iachello} is shown in Fig.~\ref{total}.

\ack
The results presented here have been obtained in  work with J. Engel, A. Escuderos, D.-L. Fang, G. Fogli, E. Lisi, A. Meroni, S. T. Petcov, V. Rodin, E. Rottuno, F. Simkovic, J. Vergados and P. Vogel \cite{Escuderos,Rod05,Ana,Fang1,Fang2,Simkovic-Vergados,Faessler1,Faessler2}. I want to thank for their collaboration.

\end{document}